# Using Gaussian Basis-Sets with Gaussian Nuclear Charge Distribution to Solve Dirac-Hartree-Fock Equation for $^{83}$Bi-Atom

Bilal K. Jasim[*], Ayad A. Al-Ani and Saad. N. Abood
Department of Physics, College of Science, Al-Nahrain University.
[*]E-mail: belalphysics28@yahoo.com.

**Abstract**

In this paper, we consider the Dirac-Hartree-Fock equations for system has many-particles. The difficulties associated with Gaussians model are likely to be more complex in relativistic Dirac-Hartree-Fock calculations. To processing these problem, we use accurate techniques. The four-component spinors will be expanded into a finite basis-set, using Gaussian basis-set type dyall.2zp to describe 4-component wave functions, in order to describe the upper and lower two components of the 4-spinors, respectively. The small component Gaussian basis functions have been generated from large component Gaussian basis functions using kinetic balance relation. The considered techniques have been applied for the heavy element $^{83}$Bi. We adopt the Gaussian charge distribution model to describe the charge of nuclei. To calculate accurate properties of the atomic levels, we used Dirac-Hartree-Fock method, which have more flexibility through Gaussian basis-set to treat relativistic quantum calculation for a system has many-particle. Our obtained results for the heavy atom (Z=83), including the total energy, energy for each spinor in atom, and expectation value of $\langle r^n \rangle$ give are good compared with relativistic Visscher treatment. This accuracy is attributed to the use of the Gaussian basis-set type Dyall to describe the four-component spinors.

**Introduction**

In many areas of physics many-particle problems are solved by generating a basis- set of suitable single-particle solutions, and then by using this basis to obtain approximate solutions for the full many-particle problem. This approach will be used to solve the Dirac-Coulomb Hamiltonian. The strategy of the Dirac-Hartree-Fock approach for calculating the electronic structure of atoms is to setup an expectation value of the Dirac-Coulomb Hamiltonian, and minimize it with respect to variations in the four-component wave function. The increasing use of all-electron four-component methodology for relativistic effects in atomic structure calculations brings with it a need for basis-set [1]. The main purpose of this work was to show that, if one starts from numerical results for atoms and fits the radial wave functions of the large components and small components of four-component spinors, one can generate in a few more steps a Gaussian basis-set which is successfully applicable to atoms such as Bi-atom. In most relativistic quantum calculations based on expansion methods, the nuclei charge distributions are described by a Gaussian charge distribution model. The Dirac equation for a single electron in the field of point charge can be solved analytically [2]. We describe the status of the problem of the electron structure of the heavy atom with nuclear charge Z=83. In relativistic calculations on heavy element consist for computational reasons of Gaussian functions and it is difficult to describe a function with a non-zero derivative at the origin. Therefore, we use relativistic basis sets to solve this problem. The four-component wave function will be expanded into a finite basis-set, by using Gaussian basis functions to describe 4-spinors. The use of Gaussian type Dyall basis-set in relativistic Dirac-Hartree-Fock calculations is likely to prove more difficult than in the corresponding non-relativistic cases. Relativistic effects are most important in heavy atoms. It will be necessary to treat these atoms species with Dirac-Hartree-Fock basis-set expansion calculations. The treatment of some relativistic effects requires an accurate description of the wave function in the inner core origin.

**Theory**

The standard Dirac-Hartree-Fock equations which contain the Coulomb interactions between the electrons are derived for system with N-electrons [3], by minimizing the expectation value of the total energy for an atom, giving:

$$E_T = \sum_a n_a \left( c\hbar \left( \int v_a(r) \left( \frac{\partial u_a(r)}{\partial r} + \frac{\kappa_a}{r} u_a(r) \right) dr - \int u_a(r) \left( \frac{\partial v_a(r)}{\partial r} - \frac{\kappa_a}{r} u_a(r) \right) dr \right) - \frac{Ze^2}{4\pi\varepsilon_0} \int \frac{1}{r} \left( u_a^2(r) + v_a^2(r) \right) dr + mc^2 \int \left( u_a^2(r) + v_a^2(r) \right) dr \right) - \sum_a \left( \frac{1}{2} n_a \frac{n_a - 1}{2j_a} \sum_{l=0}^{\infty} \frac{1}{2} (2j_a + 1) \Gamma_{j_a j_b}^l F_l(a,a) - \sum_{b \neq a} n_a \sum_{l=0}^{\infty} \frac{1}{4} (2j_b + 1) \Gamma_{j_a j_b}^l G_l(a,b) \right) + \sum_a \left( \frac{1}{2} n_a (n_a - 1) F_0(a,a) + \frac{1}{2} \sum_{b \neq a} n_a n_b F_0(a,b) \right) \quad \text{...(1)}$$

where $u(r)$ and $v(r)$ represent the radial components of the wave function. The terms inside the first summation in equation (1) represents the total energy for one electron with only one occupied shell, $F_l(a,a)$, $G_l(a,b)$ are the radial integrals and $\Gamma_{j_a j_b}^l$ represents the Clebsh-Gordan coefficient, and the terms in the second summation represent the total exchange energy. The factor $\frac{n_a - 1}{2j_a}$ after multiplying exchange energy give total exchange energy between one electron and the electrons in other shells. The terms inside the last summation represents the total Coulomb energy for an atom, the factor $\frac{1}{2} n_a (n_a - 1)$ represents the number of pairs electrons in a-shell. The 4-spinor wave function structure may be expanded in a Gaussian basis-sets as [4].

$$u_a(r) = \sum_i^N f_{\kappa p}^L(r) \xi_{ap} \quad \text{...(2)}$$

$$v_a(r) = \sum_i^N f_{\kappa q}^S(r) \eta_{aq} \quad \text{...(3)}$$

where $\xi_{ap}$ and $\eta_{aq}$ are linear variation parameters. $f_{\kappa p}^L(r)$ and $f_{\kappa q}^S(r)$ are the Gaussian basis-sets for large and small components, respectively, given by [5].

$$f^L(r) = N_L r \exp(-\zeta_L) r^2 \quad \text{...(4)}$$

$$f^S(r) = N_S r \exp(-\zeta_S) r^2 \quad \text{...(5)}$$

The factors $\zeta_L$ and $\zeta_S$ in the exponents are the only adjustable parameters of these basis functions and they are usually called the exponents of the basis function. $N_L$ and $N_S$ are normalization factors. Equation (1) has been set up for many-electron atoms. However, it is instruct to minimize it for a one-electron atom. In the one-electron limit, there is no Coulomb repulsion or exchange energy between electrons. The terms inside the first summation in equation (1) represent the total energy for one electron in an atom as:

$$E_T^1 = \sum_{a=1}^{N} n_a \left( c\hbar \left( \int v_a(r) \left( \frac{\partial u_a(r)}{\partial r} + \frac{\kappa_a}{r} u_a(r) \right) dr - \int u_a(r) \left( \frac{\partial v_a(r)}{\partial r} - \frac{\kappa_a}{r} u_a(r) \right) dr \right) + mc^2 \int \left( u_a^2(r) + v_a^2(r) \right) dr \right) \quad \text{...(6)}$$

And since one electron has only one occupied shell, so the summation disappears. Equation (6) for one electron can be rearranged slightly to become

$$E_T^1 = \int v_a(r) \left( \left( \frac{\partial u_a(r)}{\partial r} + \frac{\kappa_a}{r} u_a(r) \right) c\hbar - v_a(r)(mc^2 + V(r)) \right) dr - \int u_a(r) \left( \left( \frac{\partial v_a(r)}{\partial r} - \frac{\kappa_a}{r} v_a(r) \right) c\hbar - u_a(r)(mc^2 + V(r)) \right) dr \quad \text{...(7)}$$

where V(r) is the nuclear Coulomb potential felt by the electron. In this work we adopted the Gaussian distribution model to describe the nuclear charge. The Gaussian nuclear charge distribution is given by [2]

$$\rho_N(r_i) = Z_N \left( \frac{\eta_N}{\pi} \right)^{3/2} \exp\left( \eta_N r^2{}_{iN} \right) \quad \text{...(8)}$$

where Z is the nuclear charge and the exponent of the normalization Gaussian type function represents the nuclear charge distribution, determined by the root-mean-square radius of this distribution via the relation

$$\eta = \frac{3}{2\langle r^2 \rangle} \quad \text{...................................................} (9)$$

Where $\eta$ is the exponential parameter choosen to give a root-mean-square value. The potential V(r) in equation (7) for this charge density distribution (Gaussian model) is given by [6]

$$V(r_i) = \sum_{i=1}^{N} \int \frac{\rho_N(r_I)}{|r_i - r_I|} dr_I \quad \text{.........................} (10)$$

Where $\rho_N(r_I)$ represents the nuclear charge distribution. When the wave functions are constrained to be normalized, such that Ia, b is given by [7]

$$I_{a,b} = \int \left( u_a^*(r) u_b(r) + v_a^*(r) v_b(r) \right) dr = \delta_{a,b}$$

$$\text{...................................} (11)$$

The variation in the normalization is ΔI, as here:

$$\Delta I = 2 \int \left( \Delta u_a(r) u_b(r) + \Delta v_a(r) v_b(r) \right) dr$$

$$\text{...................................} (12)$$

If we vary u(r), while everything else remains constant, the change in energy $\Delta E_T^1$ for one electron is given by:

$$\Delta E_T^1 = \int v_a(r) \left( \left( \frac{\partial \Delta u_a(r)}{\partial r} + \frac{\kappa_a}{r} \Delta u_a(r) \right) c\hbar \right) dr -$$

$$\int \Delta u_a(r) \left( \left( \frac{\partial v_a(r)}{\partial r} - \frac{\kappa_a}{r} v_a(r) \right) c\hbar \right) dr -$$

$$2 u_a(r) \left( mc^2 + V(r) \right) dr \quad \text{...................} (13)$$

The first term in equation (13) causes some trouble, and can be solved by using integration by parts to solve varying $\Delta u_a(r)$. The right way to minimize a quantity subject to a constraint for one-electron, is to use the Lagrange multipliers method given as [8]:

$$\Delta E_T^1 - \epsilon \Delta I = 0 \quad \text{........................................} (14)$$

Substituting equation (12) and equation (13) into equation (14) and using same procedure on varying $v_a(r)$, we get the single particle Dirac equations as:

$$\frac{\partial v_a(r)}{\partial r} = \frac{\kappa_a}{r} v_a(r) - \frac{1}{c\hbar} (\epsilon - V(r) - mc^2) u_a(r) \quad \text{.................................} (15)$$

$$\frac{\partial u_a(r)}{\partial r} = -\frac{\kappa_a}{r} u_a(r) - \frac{1}{c\hbar} (\epsilon - V(r) + mc^2) v_a(r) \quad \text{.................................} (16)$$

To find the variation energy for two electrons, we used Lagrange multipliers given by [9]:

$$\Delta E_T - \sum_a \epsilon_{a,a} \Delta I_{a,a} - \sum_{a,b} \left( \epsilon_{a,b} \Delta I_{a,b} + \epsilon_{b,a} \Delta I_{b,a} \right) = 0 \quad \text{..............................} (17)$$

The variation total energy in equation (1) can be written after variations of radial integrals in direct Coulomb term $F_l(a, a)$ and exchange Coulomb term $G_l(a, b)$ to obtain Dirac-Hartree-Fock equations for the electronic structure of many-electrons atoms using Gaussian basis-set as:

$$c\hbar \left( -\frac{\partial v_a(r)}{\partial r} + \frac{\kappa_a}{r} v_a(r) \right) + \left( \epsilon_{a,a} + U_a(r) - mc^2 \right) u_a(r) + \frac{1}{2} {\sum_{b \neq a}}' \sum_{l=0}^{\infty} (2j_b + 1) \Gamma_{j_a j_b}^l \frac{1}{r} Y_l(a, b, r) u_b(r) + {\sum_{b \neq a}}' \epsilon_{a,b} u_b(r) \delta_{k_a k_b} = 0 \quad \text{...........................} (18)$$

$$c\hbar \left( -\frac{\partial u_a(r)}{\partial r} + \frac{\kappa_a}{r} u_a(r) \right) + \left( \epsilon_{a,a} + U_a(r) - mc^2 \right) v_a(r) - \frac{1}{2} {\sum_{b \neq a}}' \sum_{l=0}^{\infty} (2j_b + 1) \Gamma_{j_a j_b}^l \frac{1}{r} Y_l(a, b, r) v_b(r) - {\sum_{b \neq a}}' \epsilon_{a,b} v_b(r) \delta_{k_a k_b} = 0 \quad \text{...........................} (19)$$

The equations (18) and (19) are a pairs of Dirac-Hartree-Fock. The symbol $\sum'$ means summation over pairs. Every pair is only summed once, not twice. $U_a(r)$, represents the potential for each electron shell which differs for each electron, where $\epsilon_{a,b}$ and $\epsilon_{a,a}$ represent the diagonal and off diagonal energies, respectively. The term $Y_l(a, b, r)$ is derived from the exchange energy between the electron and the others electrons in all other shells.

**Calculation and Results**

The Dirac-Hartree-Fock radial functions of the shells occupied in the ground state, were determine for the natural atom with, Z=83. The large components $u(r)$ of these shells and the small components $v(r)$ are depicted in the figures for the atom [83]Bi. The large and small radial functions described by relativistic Gaussian basis-set of double-zeta-polarization. Fig.(1) represent the large components for all orbitals of [83]Bi atom and Fig.(2) shows the magnification of large radial functions in Fig.(1).

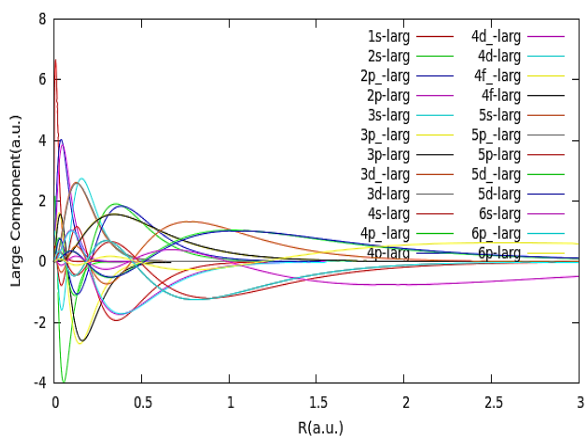

*Fig.(1): The large radial functions in atomic units against R(a.u) for all orbitals for Bi-atom using Gaussian-dyall.2zp basis-set.*

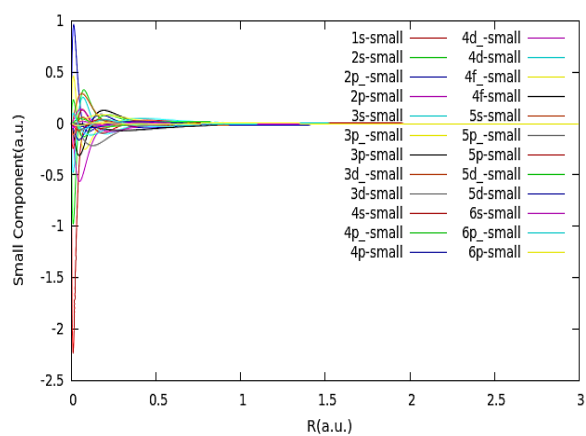

*Fig.(2): The small radial functions in atomic units against R(a.u) for all orbitals for Bi-atom using Gaussian-dyall.2zp basis-set.*

Fig.(2) shows the small component radial functions for all orbitals of $^{83}$Bi atom. It is clear that the small components radial functions are more compact and short ranged than the large component functions in Fig.(1). The effect of nuclear charge distribution on the spinor energy is notable when switching from the singular potential of point nucleus to Gaussian nucleus potential. The Gaussian nuclear potential is not different very much, most important is the effect on relative energies. The spinor energies in Dirac-Hartree-Fock level, explained in Table (1) for heavy element (Z=83) in Hartree atomic units. The results show the diffrence between two different nuclear charge distribution models. In non-relativistic theory the interaction between the electron and nuclei have traditionally been described by the simple Coulomb interaction $= -Z/r$, where r is the distance between an electron and the point nucleus with charge Z.

All the orbitals in atoms have zero amplitude at the nucleus except for s-orbital which has a cusp of the form $\exp(-\alpha r)$. In relativistic calculations, the $S_{1/2}$ spinor for Bi-element instead has a weak singularity at the nucleus as explained in Fig.(3).

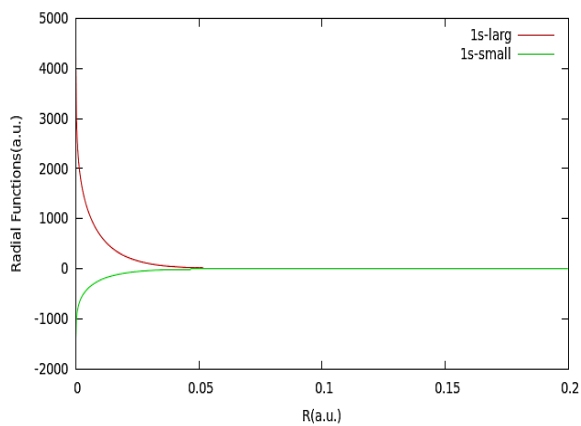

*Fig.(3): The radial $u(r)$ and $v(r)$ components in atomic units against R(a.u) for $1S_{1/2}$ orbital of Bi-atom using Slater type orbital with point model.*

In atomic calculations one expands the wave function in a large set of Gaussian basis set functions to solve the weak singularity. In this paper we adopted two models to describe the nuclear charge distribution, first model is point charge and the second is Gaussian charge model. Fig.(4) displays the radial functions $u(r)$ and $v(r)$ components for $1S_{1/2}$ of the Bi-element. The set 24s-contractive functions of Gaussian basis-set type dyall.2zp, to describe large component for the $1S_{1/2}$ spinor, and the set 20s-contractive functions to describe the small component for the $1S_{1/2}$ spinor.

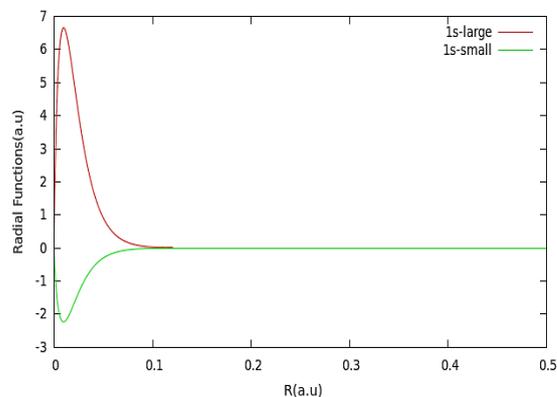

*Fig.(4): The radial $u(r)$ and $v(r)$ components in atomic units against R(a.u) for $1S_{1/2}$ orbital of Bi-atom using Gaussian-dyall.2zp basis-set with Gaussian model.*

*Table (1)*
*The relativistic spinor energy using different nuclear charge models for the heavy element (Z=83). Bismuth atom Dyall basis set is [24s 20p 14d 9f] compared with Visscher [11].*

| level | DHF Energy(a.u.) for Point model/ Our work | DHF Energy(a.u.) For Gaussian model/ Our work | Visscher /DHF-Energy (a.u.) [10] |
|---|---|---|---|
| 1s | 3352.039076 | 3349.426061 | 3352.0391 |
| 2s | 607.7970911 | 607.3929225 | 607.79709 |
| 2p- | 582.4967817 | 582.4827155 | 582.49678 |
| 2p | 497.0931648 | 497.1084527 | 497.09316 |
| 3s | 149.3877267 | 149.2941326 | 149.38773 |
| 3p- | 138.1044219 | 138.101003 | 138.10442 |
| 3p | 118.7419927 | 118.7464702 | 118.74199 |
| 3d- | 35.7578451 | 35.73385 | 35.757845 |
| 3d | 100.6180719 | 100.6222678 | 100.61807 |
| 4s | 96.55142238 | 96.55537283 | 96.551422 |
| 4p- | 30.83293247 | 30.83232473 | 30.832932 |
| 4p | 25.99901423 | 26.00045681 | 25.999014 |
| 4d- | 6.691186093 | 6.686253608 | 6.6911861 |
| 4d | 18.02529423 | 18.02656354 | 18.025294 |
| 4f- | 17.11319409 | 17.1144013 | 17.113194 |
| 4f | 4.909505587 | 4.909584734 | 4.9095056 |
| 5s | 3.976443466 | 3.976926294 | 3.9764435 |
| 5p- | 0.686847825 | 0.686193192 | 0.68684783 |
| 5p | 6.703886555 | 6.704797944 | 6.7038866 |
| 5d- | 6.49522632 | 6.496119123 | 6.4952263 |
| 5d | 1.389084536 | 1.389435764 | 1.3890845 |
| 6s | 1.270617291 | 1.270949268 | 1.2706173 |
| 6p- | 0.338421356 | 0.338483487 | 0.33842136 |
| 6p | 0.261082685 | 0.261177872 | 0.26108269 |

*Table (2)*
*Comparison of the radial expectation values <R>, <1/R> and <R²> for Gaussian model and point nucleus model of heavy element (Z=83) using Dyall basis-set with Dirac-Hartree-Fock method.*

| | Our work Point model | Our work Gaussian model | Our work Point model | Our work Gaussian model | Our work Point model | Our work Gaussian model |
|---|---|---|---|---|---|---|
| level | <R> | <R> | <1/R> | <1/R> | <R²> | <R²> |
| 1s | 0.015780291 | 0.015793731 | 103.40544 | 103.16879 | 0.000345809 | 0.000346321 |
| 2s | 0.065716919 | 0.065752034 | 25.621494 | 25.579814 | 0.00517977 | 0.00518502 |
| 2p- | 0.053955292 | 0.0539579 | 25.492566 | 25.488533 | 0.00365276 | 0.003653056 |
| 2p | 0.062962803 | 0.062961921 | 20.232288 | 20.232575 | 0.004829353 | 0.004829218 |
| 3s | 0.17067767 | 0.17074399 | 9.3992803 | 9.388873 | 0.033480075 | 0.033505608 |
| 3p- | 0.16151088 | 0.16151529 | 9.2868081 | 9.2857119 | 0.030491001 | 0.030492531 |
| 3p | 0.17783421 | 0.17783158 | 7.9154819 | 7.9155985 | 0.036755386 | 0.036754287 |
| 4s | 0.37662658 | 0.37675449 | 3.9773022 | 3.9741684 | 0.16015019 | 0.16025817 |
| 3d- | 0.15463411 | 0.15463153 | 7.7961567 | 7.7962877 | 0.028076555 | 0.028075616 |
| 3d | 0.15957911 | 0.15957658 | 7.4931312 | 7.4932489 | 0.029773422 | 0.029772471 |
| 4p- | 0.37800317 | 0.37801084 | 3.8627395 | 3.8624335 | 0.16260403 | 0.1626103 |
| 4p | 0.41101238 | 0.41100553 | 3.4089562 | 3.4090077 | 0.19196789 | 0.19196126 |
| 5s | 0.84039171 | 0.84068511 | 1.63582 | 1.6348326 | 0.79055793 | 0.79110502 |
| 4d- | 0.41504372 | 0.41503495 | 3.2095222 | 3.2095897 | 0.19885232 | 0.19884376 |
| 4d | 0.42584118 | 0.42583227 | 3.1040405 | 3.1041044 | 0.20909878 | 0.20908989 |
| 5p- | 0.89564719 | 0.89566224 | 1.5166225 | 1.5165458 | 0.90361369 | 0.90364078 |
| 5p | 0.98011627 | 0.98008874 | 1.3541795 | 1.3542096 | 1.0824889 | 1.0824228 |
| 6s | 2.2417491 | 2.2428653 | 0.57481131 | 0.57443856 | 5.7125757 | 5.7182223 |
| 4f- | 0.43646733 | 0.4364543 | 2.7683685 | 2.7684434 | 0.22683601 | 0.22682178 |
| 4f | 0.4424149 | 0.44240174 | 2.7282743 | 2.7283475 | 0.23295963 | 0.23294505 |
| 5d- | 1.2012479 | 1.2011784 | 1.090407 | 1.0904661 | 1.6655842 | 1.6653779 |
| 5d | 1.2439741 | 1.2438985 | 1.0483735 | 1.0484338 | 1.7877966 | 1.7875647 |
| 6p- | 2.7802113 | 2.7801622 | 0.45898247 | 0.45897853 | 8.907309 | 8.9068736 |
| 6p | 3.1865734 | 3.1861687 | 0.39637609 | 0.3964181 | 11.742169 | 11.738941 |

**Conclusion**

The relativistic Dirac-Hartree-Fock total energy of the ground state for Bi-atom using dyall.2zp basis sets, is -21565.70280668 a.u. with Gaussian charge model. compared with numerical calculations (visscher) is -21572.23594272 a.u. The difference in the two values is -6.5295699399 a.u. This value is not small if one takes into account. In the relativistic atomic calculations, the point charge model is not recommendable, especially, at or closer to the nuclei. This is because singularity appearance. Therefore, we adopted the Gaussian charge model combined with Gaussian basis functions to obtain accurate description for closer orbital. The total Dirac-Hartree-Fock energy for an atom depend quite a lot on the models for charge distribution.

**References**

[1] Gomes A. S., Dyall K. G., and Visscher L., "Relativistic double-zeta, triple-zeta, and quadruple-zeta basis sets for the lanthanides La–Lu". Theoretical Chemistry Accounts, 127(4), 369-381, 2010.
[2] Visser O, Aerts P. J. C., Hegarty D., and Nieuwpoort W. C., "The use of Gaussian nuclear charge distributions for the calculation of relativistic electronic wave functions using basis set expansions".

Chemical Physics Letters, 134(1), 34-38, 1987.
[3] Jasim B. K., Al-Ani A.A., and Abood S.N., "Correction Four-Component Dirac-Coulomb Using Gaussian Basis-Set and Gaussian Model Distribution for Super Heavy Element (Z=115)". Iraqi Journal of Applied Physics., 12,17-22, 2016, accepted puplished.
[4] Ishikawa Y., Quiney H. M., "On the use of an extended nucleus in Dirac–Fock Gaussian basis set calculations". International Journal of Quantum Chemistry. 32(S21), 523-532, 1987.
[5] Ishikawa Y., Baretty R., and Binning R. C., "Gaussian basis for the Dirac-Fock discrete basis expansion calculations". International Journal of Quantum Chemistry, 28(S19), 285-295, 1985.
[6] SAUE T., Fægri K., Helgaker T., Gropen O., "Principles of direct 4-component relativistic SCF: application to caesium auride"., Molecular Physics, 91(5), 937-950, 1997.
[7] Strange P., "Relativistic Quantum Mechanics: with applications in condensed matter and atomic physics". Cambridge University Press, 1998.
[8] Wilson S., Grant I. P., and Gyorffy B. L., "The effects of relativity in atoms, molecules, and the solid state". New York: Plenum Press, 1991.
[9] Slater J.C., "Quantum theory of atomic structure volume I." McGraw-Hill, New York (1960).
[10] Visscher L., Dyall K. G., "Dirac–Fock atomic electronic structure calculations using different nuclear charge distributions. Atomic Data and Nuclear Data Tables", 67(2), 207-224, 1997.

الخلاصة

في هذا البحث تم الاخذ بنظر الأعتبار معادلة ديراك هارتري فوك لنظام متعدد الالكترونات . وان الصعوبات التي ترافق نموذج جاوس ستكون اكثر تعقيدا مع حسابات ديراك هارتري فوك، ولمعالجة هذه التعقيدات سنستخدم تقنيات ذات مرونه لحل هذه التعقيدات. سنقوم بتوسيع مركبات البرم النسبية الاربعة الى مجموعة اسس محددة. أن مجموعة أسس جاوس النسبية من نوع dyall. 2zp استخدمت لوصف دوال الموجه النسبية ذات المركبات الاربعة وأن مركبات دوال أسس جاوس النسبيه الصغيره يمكن توليدها من مركبات دوال أسس جاوس الكبيره باستخدام دالة التوازن الحركية. وتم تبني نموذج توزيع كاووس لوصف توزيع شحنة النواة وطبقت هذه التقنية على عنصر ذري ثقيل وهو (Bi, Z=83). للحصول على حسابات اكثر دقة للخصائص الذريه. تم استخدام طريقة ديراك هارتري فوك مع مجموعه أسس جاوس النسبية لمعالجة الحسابات الكمية النسبية لنظام متعدد الذرات. ان النتائج التي حصلنا عليها كانت اكثر دقه من نتائج معالجة Visscher النسبية ويعزى ذلك لاستخدامنا مجموعة أسس جاوس النسبية من نوع Dyall لوصف المركبات الاربعة النسبيه ونموذج توزيع جاوس.